# Artificial Intelligence in Pharmaceutical Development:
## Building Rigorous, Decision-Ready Evidence

Xin Luo[1*], Margaret Gamalo[1], Achilleas Livieratos[2], Maria Kudela[1], Qilu Yu[3], Yunxun Wang[4], Junjing Lin[5], Yuxi Zhao[1], Allshine Chen[1]

[1]*Pfizer Inc., 500 Arcola Road, Collegeville, PA 19426, USA*
[2]*Independent Researcher, 152 38 Athens, Greece*
[3]*Office of Clinical and Regulatory Affairs, National Center for Complementary and Integrative Health, National Institutes of Health, 6707 Democracy Boulevard, Suite 401, Bethesda, MD 20892, USA*
[4]*Bayer U.S. LLC, 100 Bayer Boulevard, Whippany, NJ 07981, USA*
[5]*Takeda Development Center Americas, Inc., 500 Kendall Street, Cambridge, MA 02142, USA*

****Corresponding author:*** xin.luo@pfizer.com

**ABSTRACT**

Artificial intelligence (AI) and machine learning (ML) are reshaping the pharmaceutical lifecycle, from discovery and clinical development through manufacturing and regulatory review to safety surveillance and health economics, creating evidentiary challenges that predictive performance alone cannot address. This review examines the inferential, causal, validation, and documentation requirements for AI-enabled evidence generation in high-stakes settings and argues that they must be built into workflows. We outline the evidentiary functions that AI-derived outputs must satisfy before supporting consequential decisions: estimand definition, fit-for-purpose data collection, bias control, uncertainty quantification, validation, interpretability, and decision-grade documentation. Because AI both depends on and accelerates large-scale, non-randomized data collection, the integration of randomized controlled trials (RCTs) with real-world data (RWD) is where its evidentiary risks concentrate; we therefore present a five-step causal roadmap for that integration, in which external data supplement but never replace a concurrent control, and characterize the safeguards required across development, pharmacovigilance, regulation, and health technology assessment. Priority frontiers answer recurring implementation problems: conformal prediction for overconfident predictions, randomization-based inference for scarce data, and federated causal inference and privacy-preserving generative modeling for data that cannot be pooled. Evidence generation depends not on competition between RCTs and RWD, or between statistics and AI/ML, but on their principled synthesis under causal and inferential frameworks, with statistical expertise central to decision-ready evidence.

## 1. Introduction

Artificial intelligence (AI) and machine learning (ML) are transforming the pharmaceutical development lifecycle [1–3]: from discovery and manufacturing through clinical trials, regulatory review, post-authorization safety surveillance, health economics and outcomes research (HEOR), and market access, they create new capabilities and obligations. A recent perspective addressed AI in Medical Affairs alone [4]; this review concentrates on clinical development through commercialization, where inferential and decision-making demands are most acute.

Regulators have responded: the European Medicines Agency (EMA) 2024 Reflection Paper [5], the U.S. Food and Drug Administration (FDA) 2025 guidance [6], and the agencies' joint 2026 Guiding Principles of Good AI Practice in Drug Development [7] codify standards demanding validation, transparent documentation, uncertainty quantification, and bias assessment beyond performance metrics.

Our central argument is that AI-enabled evidence generation requires inferential discipline beyond prediction. Predictive accuracy alone is not scientific evidence: in high-stakes settings, AI outputs must be tied to explicit questions, built on data collected for the purpose, evaluated against plausible biases, accompanied by calibrated uncertainty, and documented to withstand scrutiny. This demands statistical thinking in fit-for-purpose data collection, estimation, testing, causal identification, and uncertainty propagation.

A complementary perspective [8] holds that clinical evidence generation depends on principled integration of randomized controlled trials (RCTs) with real-world data (RWD) under causal inference frameworks rather than on methodological competition. The focus on RWD is not incidental: AI both depends on and accelerates the collection of large, non-randomized datasets, so RCT–RWD integration is where its evidentiary risks concentrate. Calls for safeguards range from roadmaps for responsible ML in health care [9] and critical questions on transparency, replicability, and effectiveness [10] to reporting standards for AI-enabled trials and prediction models [11, 12]. MetaMind [13] and van der Schaar's multi-agent framework [14] show AI extending evidence synthesis and trial monitoring under statistical safeguards.

## 2. Core Evidentiary Functions for AI in Pharmaceutical Development

Beyond performing well on held-out data, AI systems supporting consequential decisions must address clearly framed questions, rest on fit-for-purpose data, operate under explicit assumptions, quantify uncertainty, withstand bias diagnostics, and remain interpretable and documentable. Table 1 summarizes these functions and the expertise they require; Figure 1 maps them onto the lifecycle.

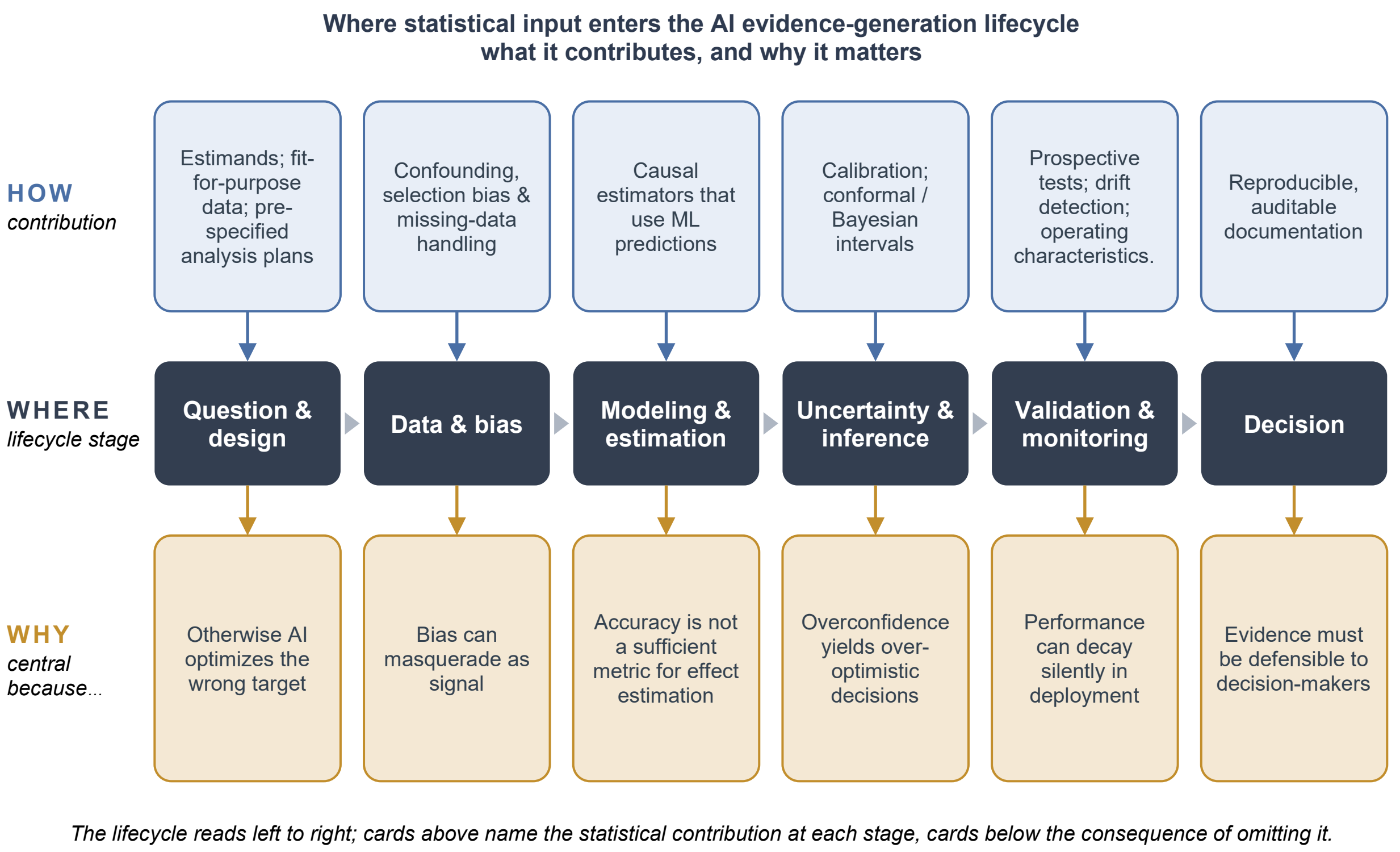

**Figure 1.** Statistical input across the AI evidence-generation lifecycle: each stage (center), its statistical contribution (above), and the failure mode of omitting it (below).

## 2.1 Methodological Rigor, Bias Control, and Inferential Performance

Rigorous AI use begins before model fitting: experimental design, estimand definition, fit-for-purpose data collection, and pre-specified analysis plans determine whether outputs are evidence rather than artifacts; no model sophistication compensates for underpowered or misaligned studies. In early-phase oncology, for example, the continual reassessment method replaces rule-based escalation with a pre-specified dose–toxicity model updated as data accrue, concentrating patients near the maximum tolerated dose while limiting exposure to toxic doses [15].

AI development emphasizes model construction and performance optimization, whereas pharmaceutical evidence generation also requires formal attention to estimand alignment, inferential operating characteristics, multiplicity control, assumption checking, and reproducibility. A model that performs well on held-out data may still yield biased or fragile conclusions, and poorly calibrated or overconfident outputs distort efficacy estimates and compromise safety surveillance.

Bias identification and control are central to any AI application supporting causal or comparative claims. Bias arises not only from algorithmic behavior but from the data-generating process: confounding, selection, measurement error, outcome misclassification, and nonrepresentative sampling. Propensity score adjustment, causal diagrams, sensitivity analyses, and structured subgroup assessment make these risks explicit and auditable; analyses that misclassify the pre-treatment interval (immortal time bias) can make even ineffective drugs appear protective, and aligning time zero within a target trial emulation removes the artifact [16].

Outputs must also be interpretable: reviews catalog how explainable machine learning applies in clinical research and where it fails in high-stakes decisions [17], and decisions about treatment effects, safety, and reimbursement require calibrated uncertainty. Epistemic uncertainty, from limited data or misspecification, shrinks with more data, whereas aleatoric uncertainty, the irreducible variability of biology and measurement, bounds the precision any model can achieve and must be communicated rather than engineered away. A predictor with calibrated uncertainty can defer to human judgment when its confidence is low, which turns prediction into decision-ready output.

### 2.2 Decision Frameworks, Documentation, and Complex Data Structures

Regulators, health technology assessment (HTA) bodies, and internal governance boards set the expectations for credible AI evidence in high-stakes decisions, and complex or missing data are where meeting them is hardest. Regulatory frameworks are the most codified instance: the FDA [6] and EMA [5] require risk-based assessment of AI used in regulatory decision-making, emphasizing credibility, validation, oversight, and monitoring [7], with validation plans and documentation that align a model's context of use with the claims made. The first FDA-authorized autonomous diagnostic AI was cleared on a prospective pivotal trial with pre-specified sensitivity and specificity targets, not retrospective accuracy [18].

These expectations are most demanding where data are least well behaved. RWD from electronic health records, claims, and registries have complex structure, high missingness, and non-random collection, so missing-data mechanisms and provenance assessment are core evidentiary functions. The International Council for Harmonisation (ICH) E9(R1) estimands framework requires pre-specified handling of intercurrent events and missing data, which matters acutely in Alzheimer's disease trials, where progression-related dropout can bias a naive treatment effect [19].

### 2.3 Translation, Ethical Oversight, and Patient Safety

AI methods are useful only when translated into clinical and operational decisions, which requires attention to workflow, patient safety, interpretability for end users, and the consequences of failure: a model that ignores site-level variability in drug administration, or a surrogate endpoint misaligned with regulatory expectations, is impressive yet unusable. Statisticians bridge both domains by quantifying how assumption violations affect clinical conclusions and validating outputs for statistical rigor and clinical relevance alike.

Ethical oversight is inseparable from quantitative rigor. Fairness, patient protection, and acceptable risk must be operationalized through measurable error rates, subgroup performance, calibration, and surveillance, from trial stopping rules to audits of deployed systems. Pharmacovigilance (PV) makes the trade-off explicit: disproportionality methods applied to the Vaccine Adverse Event Reporting System (VAERS) and the FDA Adverse Event Reporting System (FAERS) accept many false-positive product–event pairs so that rare harms surface early [20], and because these data are shaped by under-reporting, stimulated reporting, and missing exposure denominators, the screening threshold is an ethical choice about which errors patients will bear.

**Table 1. Core evidentiary functions for AI implementation in pharmaceutical development**

| Evidentiary function | Why it matters | Failure mode if neglected | Typical contributors |
|---|---|---|---|
| Experimental design, estimand definition, and fit-for-purpose data collection | Aligns the analysis with the question and decision context, and the data with the intended inference | Strong performance on the wrong question, or pipelines built on unrepresentative data | Statistics, clinical science, regulatory science, data engineering |
| Bias identification and causal validity | Needed for comparative claims, external controls, transportability, and RWE | Confounded, nontransportable, or missingness-driven conclusions presented as causal evidence | Statistics, epidemiology, clinical experts |
| Uncertainty quantification and interpretability | Supports risk-aware decisions, human oversight, and regulatory review | Overconfident predictions or opaque outputs used in high-stakes settings | Statistics, machine learning experts, clinicians, regulators |
| Validation and decision-grade documentation | Establishes credibility for a context of use and preserves auditability | An impressive model that cannot support a regulatory, reimbursement, or governance decision | Statistics, regulatory affairs, quality, machine learning experts |
| Clinical translation and lifecycle monitoring | Keeps outputs useful, safe, and interpretable in deployment | Drift, subgroup harm, or workflow failure discovered only after deployment | Clinical teams, statistics, pharmacovigilance, machine learning experts |

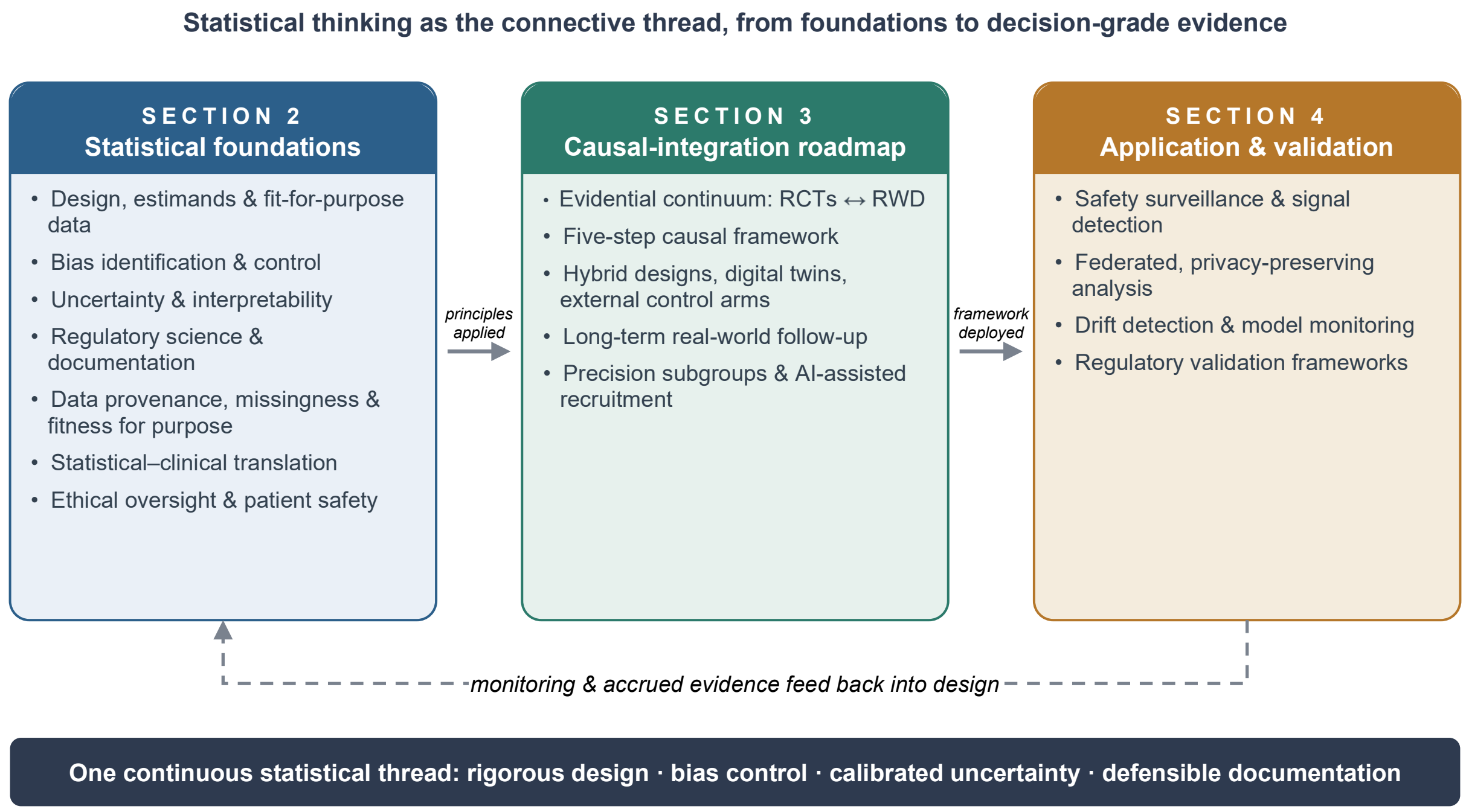

**Figure 2.** Statistical thinking as the connective thread: the Section 2 foundations are operationalized in the Section 3 roadmap and applied in Section 4, with monitoring feeding back into design.

# 3. A Causal Roadmap for Integrating RCTs and RWD

## 3.1 The Evidential Continuum and the Five-Step Framework

Randomized controlled trials are the gold standard for causal treatment effects, providing internal validity through randomization [21], but they are expensive, slow, and limited in external validity by eligibility criteria that exclude elderly patients, those with comorbidities, and underrepresented groups. RWD offer complementary breadth; in 2019 to 2021, real-world evidence (RWE) contributed to approximately 31% of FDA approvals of new drugs and biologics [22, 23]. The core insight [8] is that evidence generation depends not on RCTs versus RWD, or statistics versus AI/ML, but on their principled synthesis: RCTs establish internal validity, RWD extends generalizability, AI/ML expands analytical capacity, and statistical causal inference supplies the scaffolding that makes the integration defensible [24]. The roadmap has five steps (Figure 2).

**Step 1 Define the causal estimand.** The scientific question, formalized as an estimand under ICH E9(R1), must be specified before any data are examined. In the statin–cancer example below, the estimand is the 10-year difference in cancer-free survival between initiating and not initiating statins in statin-eligible adults, under an intention-to-treat contrast; a model predicting cancer risk among current statin users answers a different question.

**Step 2 Align data sources and articulate identification assumptions.** Data provenance and fitness for purpose must be assessed explicitly, and potential biases (confounding, selection, temporal misalignment, immortal time, outcome misclassification) identified and their plausibility evaluated with causal diagrams informed by subject-matter expertise [25], making explicit the assumptions needed to interpret associations as causal effects.

Target trial emulation makes Step 2 a design discipline: the analyst writes the protocol of the ideal randomized trial and aligns the observational data to each element, from eligibility and time zero to outcome and causal contrast [26, 27]. Dickerman and colleagues applied it to the claim that statins prevent cancer in primary care records of 733,804 adults [28]. Classifying individuals by eventual duration of statin use gave a hazard ratio of 0.22 for total cancer; the artifact is misaligned time zero, since a long-term user must survive cancer-free long enough to become one [16]. With eligibility, assignment, and follow-up aligned at one time zero, the 10-year cancer-free survival difference was −0.5% (95% confidence interval [CI]: −1.0% to 0.0%) and the hazard ratio 1.02 (95% CI: 0.99 to 1.05); no Step 3 estimator could have repaired the design.

**Step 3 Select estimation methods aligned with the causal estimand.** Doubly robust estimators such as targeted maximum likelihood estimation (TMLE) and augmented inverse prob-ability weighting (AIPW) let flexible AI/ML algorithms estimate nuisance functions while retaining root-n valid inference [29]; double machine learning with sample splitting removes regularization bias in high-dimensional settings [30], and Dragonnet [31], TARNet [32], and causal transformers [33] learn balanced representations and treatment effects jointly. In the statin emulation, the treatment-initiation and censoring models are the nuisance functions: inside a doubly robust estimator they can be fitted by gradient boosting and the estimate remains valid if either model is correct, whereas plugging ML predictions into an outcome regression inherits their regularization bias [30].

**Step 4 Assess robustness of conclusions to assumption violations.** Sensitivity analyses quantify how violations of identification and modeling assumptions would alter conclusions, complemented by E-values [34] and negative control outcomes and exposures [35]. An observed hazard ratio of 1.5, for example, has an E-value of 2.37: an unmeasured confounder would have to be associated with both treatment and outcome by a risk ratio of at least 2.37 to explain the association away [34], and an association with an outcome the treatment cannot plausibly affect signals residual confounding [35].

**Step 5 Contextualize the evidence within the decision problem.** A validated causal estimate must be situated within the totality of evidence, assessed for transportability, framed in benefit–risk terms, and accompanied by its residual uncertainties. The myocarditis case in Section 4.1 is Step 5 in practice: a rate ratio near 10 became actionable only when set against the Covid-19 hospitalizations and deaths averted per million doses.

## 3.2 Hybrid Trial Designs, Digital Twins, and External Control Arms

In rare diseases and oncology, ethical and feasibility constraints often preclude a fully randomized control arm. External control arms (ECAs), real-world cohorts matched to the trial arm on prognostic characteristics, address this; AI enters at cohort assembly, harmonization, imputation, outcome extraction, and matching. Replacing randomization entirely with synthetic or real-world comparators risks uncontrolled confounding and removes what borrowing depends on: a concurrent control. Without a randomized arm there is nothing against which to measure, and down-weight the discrepancy of the external data, so hybrid designs that retain a small randomized arm and borrow against it are the defensible option.

Oncology external controls show what any comparator, AI-curated or not, must demonstrate. Avelumab received accelerated approval for metastatic Merkel cell carcinoma in 2017 on a single-arm trial of 88 chemotherapy-refractory patients, one third of whom responded [36]. The sponsor's chart-review cohort found only 20 second-line chemotherapy patients among 686, with a 28.6% response rate and 1.7-month median response duration [37]; FDA treated it as exploratory, made no formal comparison, and acknowledged its selection bias [36]: without a concurrent control, approval could rest only on response magnitude and durability. Across seven oncology external control submissions reviewed in 2014 to 2021, selection bias and confounding were the most common regulator and HTA critiques [38], and FDA's 2023 draft guidance now expects pre-specified data sources and eligibility, an immortal-time-free index date, comparable outcome ascertainment, and patient-level control-arm data [39]. Every requirement concerns the design and provenance of the comparison, not matching quality, and an AI-assembled cohort must meet each one auditably.

Digital twins, computational models of individual patients built from longitudinal data, simulate counterfactual trajectories. The Prognostic Covariate Adjustment (PROCOVA) platform, which uses digital twin predictions as prognostic covariates in analysis of covariance (ANCOVA), received an EMA qualification opinion in 2022 [40] and falls within FDA's 2023 covariate-adjustment guidance [41]. The borrowing in hybrid designs is itself statistical rather than AI: power, commensurate, and meta-analytic predictive priors downweight external data through a pre-specified, interpretable function of the observed discrepancy with the randomized arm, and frequentist test-then-pool and selective-borrowing methods provide analogous protection [42]; the statistician fixes the degree of skepticism in advance, and every variant presupposes a control, since external data can sharpen a randomized comparison but cannot create one. AI enters upstream, in assembling or synthesizing the external data, which is where opacity must be audited.

PROCOVA shows why this AI support strengthens rather than replaces randomization. A prognostic model trained on historical data independent of the trial predicts each participant's outcome under control; the model is fixed and its prediction pre-specified as an ANCOVA covariate [42]. Because treatment remains randomized, the estimator keeps the type I error control and unbiasedness of ordinary ANCOVA whatever the model's quality; validity does not rest on exchangeability between the historical and trial populations, but the precision gain does, and a model transported to a dissimilar population forfeits it. The Committee for Medicinal Products for Human Use accordingly qualified PROCOVA for continuous endpoints only, declined to qualify any model-building procedure, insisted on the external validity (transportability) of the historical data, and required that sample size reductions (typically 10% to 30%) account for uncertainty in the assumed gain [40, 42]. The guarantee comes from randomization and pre-specification, not from the model; for binary and time-to-event endpoints, adjustment changes the estimand and the same protection requires model-robust marginal estimators [41].

### 3.3 Long-Term Follow-Up and Subgroup Discovery

Trials are powered for short-term or intermediate endpoints, leaving durability, late adverse effects, and long-term cost-effectiveness unanswered. Linking trial cohorts to registries or claims addresses these gaps if informative censoring, treatment switching, and time-varying confounding are handled explicitly, and Germany's Federal Joint Committee [43], the UK's National Institute for Health and Care Excellence (NICE) [44], and the U.S. Institute for Clinical and Economic Review [45] prioritize long-term RWE in reimbursement decisions.

AI/ML methods expand the capacity to discover treatment effect heterogeneity; causal forests, for example, provide non-parametric individual-level estimates with formal guarantees [46]. AI-identified subgroups risk spurious findings through implicit multiple testing and post-hoc selection, so statisticians apply bootstrap stability testing, cross-validation, and Shapley Additive Explanations (SHAP)-based plausibility checks, and require confirmatory analysis.

AI-assisted design can also introduce active learning, with RWD-derived prognostic models steering recruitment toward the participants most informative for the causal objective, increasing precision in rare diseases where recruitment is the binding constraint [47]. FDA's 2026 draft guidance on Bayesian methods endorses borrowing from RWD and earlier trials to construct informative priors that supplement, not replace, a concurrent control [48].

# 4. AI in Pharmacovigilance and Regulatory Affairs

## 4.1 Safety Surveillance and Federated Signal Detection

Pharmacovigilance faces a structural limit: no single organization has enough adverse event data to detect rare but serious signals, and privacy regulations prevent pooling. Federated learning trains models jointly without pooling individual-level data, each institution contributing only model parameter updates [49], so the federated model draws on the collective power of all institutions and improves sensitivity for rare adverse events [50].

Statisticians safeguard these systems through secure aggregation protocols, differential privacy against re-identification from gradient updates, validation of federated against centralized performance, and drift detection as populations, coding practices, or treatment patterns evolve.

As AI-based natural language processing (NLP) and classification models increasingly triage adverse event reports [51], interpretability becomes a regulatory prerequisite. Explainable AI (XAI) techniques [52], particularly SHAP values and Local Interpretable Model-agnostic Explanations (LIME), give PV teams a rationale to act on AI-generated signals, and FDA and EMA treat interpretability as a condition of acceptance.

Myocarditis after mRNA Covid-19 vaccination shows the chain from signal to decision. By June 2021, VAERS had received 1,226 myocarditis reports; screening against expected background rates [20] located a cluster in young males after the second dose [53]. Centers for Disease Control and Prevention (CDC) physicians reviewed each report against the CDC case definition, and 1,626 through August 2021 met it, peaking at 105.9 per million second doses in males aged 16 to 17 [54]. The Vaccine Safety Datalink, sequentially monitoring 11.8 million doses, estimated a rate ratio of 9.83 in the week after vaccination [55]. The Advisory Committee on Immunization Practices weighed these risks against Covid-19 hospitalizations and deaths averted and concluded that benefits clearly outweighed risks; FDA and CDC updated labeling and clinical guidance [53, 56]. NLP triage accelerates the first link in this chain but does not replace adjudication, active surveillance, or benefit–risk contextualization.

### 4.2 Regulatory Validation Frameworks for AI

The EMA's 2024 Reflection Paper [5] sets out a risk-stratified framework for AI in the medicinal product lifecycle, FDA's 2025 guidance [6] introduces analogous expectations for documentation, performance monitoring, and pre-specified validation criteria, and the joint FDA–EMA principles of January 2026 consolidate these into ten principles [7].

Statisticians in regulatory affairs translate these principles into operational criteria: which performance metrics a model must demonstrate, on what validation dataset, within which pre-specified subgroups (site, demographic group, disease severity), and against what acceptance criteria before it may support a submission; they author validation plans and lead prospective monitoring.

A point prediction without a prediction interval for an individual forecast, or a confidence or credible interval for a population-level estimate, is insufficient for high-stakes decisions (Section 6.1).

## 5. AI in Health Economics, Outcomes Research, and Market Access

HEOR and market access concentrate the Section 2 competencies on one question: whether HTA agencies and payers will accept AI-derived RWE for reimbursement. Demonstrating value with observational data vulnerable to confounding and selection depends on causal anchoring, transparent documentation, and uncertainty quantification, an evidentiary design problem rather than a modeling problem.

The NICE real-world evidence framework turns this design problem into an explicit protocol [44]. It delimits the admissible uses of RWE, from populating economic models to constructing external controls; requires fitness for purpose to be documented before analysis through the Data Suitability Assessment Tool (provenance, quality, relevance); and requires comparative-effect studies to emulate the ideal randomized trial, with a pre-registered protocol and with confounding, informative censoring, missing data, and measurement error addressed through quantitative bias analyses. Each element maps onto a Table 1 function, and an AI pipeline must leave an auditable record at each.

Transportability analysis, generalizing RCT treatment effects to the broader RWD populations, is among the most policy-relevant AI/ML applications in HEOR, complemented by target trial emulation for long-term questions [26].

Pragmatic trials embed randomization within routine care, though looser operational controls introduce heterogeneity [57]. ADAPTABLE illustrates the promise and obligations [58]: through the Patient-Centered Clinical Research Network (PCORnet), it identified eligible patients from electronic health records at 40 centers, randomized 15,076 patients to 81 mg or 325 mg of daily aspirin, and ascertained outcomes from records and claims; over a median 26.2 months, the primary outcome occurred in 7.28% versus 7.51% (hazard ratio 1.02; 95% CI: 0.91 to 1.14). The price of pragmatism was also exposed: 41.6% of patients assigned to 325 mg switched to 81 mg, so intention-to-treat estimates the effect of assigning a dose rather than taking it, and the per-protocol question needs the adjustment for time-varying adherence that observational analyses require [57]. AI-assisted screening and outcome extraction can scale such trials but inherit these obligations.

Statisticians lead HTA-grade analytics: pre-specified RWE analysis plans, Bayesian decision-analytic frameworks that propagate uncertainty through cost-effectiveness models, and economic simulations whose parameters reflect the inferential uncertainty of the underlying estimates.

## 6. Priority Methodological Frontiers

The frontiers below each answer a problem that recurs in the preceding sections, and they differ in maturity: conformal prediction addresses overconfident black-box predictions and is theoretically mature, with regulatory-facing applications emerging; randomization-based inference addresses small samples and is classical in origin but only recently combined with selective borrowing; federated causal inference and privacy-preserving generative modeling address data that cannot be pooled and are the least mature.

### 6.1 Calibrated Uncertainty for AI-Derived Estimates

When the goal is to inform treatment or policy decisions, the question is causal, and models optimized for outcome accuracy can be well calibrated yet causally misleading; causal AI/ML methods anchor estimation to defined estimands, so that intervals quantify the precision of a genuine treatment effect, and conformal prediction complements this with distribution-free, finite-sample valid prediction intervals for any model, including black boxes [59].

### 6.2 Inference Under Data Scarcity and Across Institutions

Rare diseases, pediatric indications, and early-phase oncology operate under data constraints where asymptotic approximations fail and naive machine learning overfits. Randomization-based inference [60], which derives validity from the randomization distribution itself, preserves it in data-limited trials; Zhu and colleagues combine it with conformal selective borrowing in hybrid controlled trials, retaining exact type I error control while borrowing only external controls that resemble the randomized ones, a judgment possible only because a randomized control exists [60].

Federated learning (Section 4.1) lets institutions train models jointly while patient-level data never leave home: the EXAM model for predicting oxygen requirements in Covid-19 was trained across 20 institutions on five continents without sharing patient data and generalized better than any single-site model [61]. Generative modeling complements it: diffusion and related frameworks can synthesize non-identifiable datasets, though their fidelity and privacy guarantees must be validated rather than assumed [62]. The safeguards of Section 4.1 apply unchanged.

### 6.3 Prospective Validation of Real-World Evidence

Extending principled causal inference to AI-driven RWE requires scaffolding that disciplines AI-detected associations into defensible causal claims. Many identification assumptions, particularly exchangeability in observational comparisons, cannot be verified retrospectively; embedding small, prospective randomized sub-studies as calibration anchors bounds assumption violations more directly than sensitivity analysis alone. When pooling is infeasible, federated causal inference [63] extends the analysis without centralized data access.

## 7. Discussion: Interdisciplinary Governance and Quantitative Leadership

The challenge of AI integration is not primarily technological: methods can be automated, but judgment about estimands, identification assumptions, data provenance, and residual uncertainty cannot, and that judgment is the binding constraint. AI outputs must be tied to explicit estimands, built on fit-for-purpose data, screened for bias, accompanied by calibrated uncertainty, validated for their context of use, monitored for drift and subgroup harms, and documented to withstand scrutiny; each is a Section 2 function, and together they call for quantitative leadership and genuine interdisciplinarity.

Quantitative expertise must therefore be engaged from question formulation and data-collection planning onward, not only in late-stage validation; pre-specifying analysis plans, causal estimands, and validation criteria before examining data is as important for AI-enabled RWE as for randomized trials.

## 8. Conclusions

AI is opening a transformative chapter in pharmaceutical development, but realizing that potential depends on principled integration of AI's flexibility with causal, inferential, and validation frameworks. The decisive question is not whether AI can generate signal but whether that signal can become evidence that is decision-ready, transportable, and accountable. This is where statistical expertise matters most, in defining estimands, securing fit-for-purpose data, characterizing bias, quantifying uncertainty, designing validation, and clarifying inferential limits, within interdisciplinary teams combining quantitative rigor with clinical, computational, and regulatory expertise.

Statistical thinking remains central in any high-stakes decision, regulated or not, where causal claims, uncertainty, and operating characteristics matter directly. Done well, this repositioning will strengthen the evidentiary foundations of development decisions and translate AI's promise into treatments that reach patients faster, more safely, and with greater confidence in their real-world benefit.

## Funding Statement

This review received no specific funding from any agency in the public, commercial, or not-for-profit sectors.

## Conflict of Interest Statement

The authors declare no competing financial interests. XL, MG, MK, YZ, and AC are employees of Pfizer Inc.; AL is an independent researcher; QY is an employee of the National Center for Complementary and Integrative Health, National Institutes of Health; YW is an employee of Bayer U.S. LLC; and JL is an employee of Takeda Development Center Americas, Inc.. Pharmaceutical company employees may hold stock or stock options in their employers.

## Author Contributions

XL conceived the review, led the literature synthesis, and drafted and revised the manuscript. MG, MK, YZ, AC, AL, QY, YW, and JL contributed to the conception, scope, and content, and reviewed the manuscript critically. All authors approved the final version and agree to be accountable for all aspects of the work.